\documentclass[twocolumn,showpacs,amsmath,amssymb,a4paper,flushleft]{article}
\usepackage{graphicx}
\usepackage{dcolumn}
\usepackage{bm}
\usepackage{cite}

\newcommand{\vect}[1]{\mbox{\boldmath $#1$}} 

\begin{document}
\renewcommand{\thefootnote}{\fnsymbol{footnote}}
\twocolumn[
{\LARGE Supramolecule Structure for 
Amphiphilic Molecule \\
by Dissipative Particle Dynamics Simulation}

\vspace{2ex}
Hiroaki NAKAMURA\footnotemark

\vspace{2ex}
{\small \it Theory and Computer Simulation Center
/ The Graduate University for Advanced Studies,
National Institute for Fusion Science,
322-6 Oroshi-cho, Toki, Gifu 509-5292, JAPAN}
\vspace{3ex}
({\it Received \today})\\
]
\footnotetext{e-mail: nakamura@tcsc.nifs.ac.jp}

\noindent
{\bf Meso-scale simulation of structure formation for AB-dimers in solution 
W monomers
was performed by dissipative particle dynamics (DPD) algorithm. 
As a simulation model, modified Jury Model 
was adopted~[Jury, S. {\it \bf et al}. ``Simulation of amphiphilic mesophases using 
dissipative particle dynamics," 
{\bf \it Phys. Chem. Chem. Phys.} 
{\bf 1} (1999) 2051--2056], which  represents mechanics of self-assembly  for 
surfactant hexaethylene glycol dodecyl ether (${\rm C}_{12}{\rm E}_6$) and 
water(${\rm H}_2$O).
The same phase diagram as Jury's result was obtained.
We also found that it takes a longer time to form the hexagonal  phase (${\rm H}_1$) 
than to form the lamellar phase (${\rm L}_{\alpha} )$.
}

\vspace*{2ex}
\noindent
{\it \small Keywords}{\small : Amphiphilic molecule, Nonionic surfactant, 
Dissipative particle dynamics simulation,
Self assembly, Structure formation, Phase diagram}

\vspace{3ex}
\noindent
\section*{\normalsize INTRODUCTION}

Dissipative particle dynamics (DPD) simulation has become
one of the most powerful algorithm for soft-matter 
research\cite{92Hooger,97Groot,98Groot,01Groot}.
In DPD simulation, the conservative interaction potentials
are soft-repulsive, which make simulation of long-time phenomena possible.
By the way, DPD algorithm might be considered one of 
the coarse-grained method of molecular dynamics(MD) simulation.
Some information of interaction potential between particles
are neglected and simplified. We, therefore, must pick up the dominant
interaction potential for the mesoscopic structure formation.
Because we don't have enough experimental data for the interaction potentials,
it becomes a difficult problem  of DPD simulation 
how we define interaction parameters. 
In 1999, Jury {\it et al.} succeeded, by an empirical method, 
the DPD simulation of the smectic mesophase 
for a simple amphiphilic molecule system with water solvent \cite{99Jury}.
They showed that their minimal model (we call it Jury model), which is composed of rigid AB
dimers in solution W monomers,
is very proper to present phase diagram of surfactant hexaethylene
glycol dodecyl ether (${\rm C}_{12}{\rm E}_6$) and water 
(${\rm H}_2 {\rm O}$)\cite{99Jury,83Mitchell}.
In this paper, we reveal processes of   self-organization of one smectic mesophase
by modified Jury model which has such a  difference from the original Jury model 
that attractive harmonic oscillator potential is used as a  covalent bond potential
between A and B particles in the same molecule. 
Details of simulation method are explained in the next section.
In the last section, we will show such a interesting property  of the system
that it take a longer time to form the hexagonal  phase (${\rm H}_1$) 
than the lamellar phase $({\rm L}_{\alpha})$ which is more symmetric than 
${\rm H}_1$ phase.
It is considered intuitively that 
it takes a longer time to form the higher ordered phase than
to form the lower ordered phase.
However, this simulation shows that
it takes a longer time to form the lower ordered phase than
to form the higher ordered phase contrary to the intuitive consideration.

\section*{\normalsize SIMULATION METHOD}

\subsection*{\normalsize DPD Algorithm}
First of all, we express the DPD model and algorithm\cite{97Groot,99Jury}.
According to ordinarily DPD model, 
all atoms are coarse-grained to particles whose mass are the
same one.
We define the total number of particles as $N.$
The position and velocity vectors of particle $i, (i=1,\cdots, N) ,$
are indicated by $\vect{r}_i$  and $\vect{v}_i$, respectively. 
The particle $i$ moves according to the following equation of motion,
where all physical quantities are made dimensionless to 
handle easily in actual simulation.
\begin{eqnarray}
\frac{d \vect{r}_{i} }{d t} &=& \vect{v}_i ,      \label{eq1} \\
\frac{d \vect{v}_{i} }{d t} &=& \sum_{j (\ne i)}^{N} \vect{F}_{ij},   
\label{eq2}
\end{eqnarray}
where a particle $i$ interacts with the other  particle $j$ 
by  $\vect{F}_{ij}$ is the total force
\begin{equation}
\vect{F}_{ij} =
\vect{F}_{i j}^{\rm C} + \vect{F}_{i j}^{\rm R} +\vect{F}_{i j}^{\rm D}  
                        +\vect{F}_{i j}^{\rm B}  . \label{eq3}
\end{equation}
In Eq.~\ref{eq3},
$\vect{F}^{\rm C}_{ij}$ is a conservative force deriving from a potential 
exerted on particle $i$ by the $j-$particle, $\vect{F}^{\rm D}_{ij}$ and 
$\vect{F}^{\rm R}_{ij}$ are the dissipative and random forces between 
particle $i$ and $j$, respectively. Furthermore, neighboring particles 
on the same amphiphilic molecule are bound by the  bond-stretching force 
$\vect{F}^{\rm B}_{ij}$. 

The conservative force $\vect{F}^{\rm C}$ has the following form:
\begin{equation}
\vect{F}^{\rm C}_{ij} = - \nabla_i \phi_{ij} , \label{eq.fc}
\end{equation}
where $\nabla_i \equiv \partial/\partial \vect{r}_i $. 
For computational convenience, we adopted that the cut-off length as 
the unit of length.
It is assumed that the conservative force $\vect{F}^{\rm C}$ are truncated 
at this radius.
Following this assumption, 
the two-point potential $\phi_{ij}$ in Eq.~\ref{eq.fc} is defined as follows:
\begin{equation}
\phi_{ij} \equiv \phi(r_{ij}) =  
\frac{1}{2} a_{ij} \left( r_{ij} - 1 \right)^2 
                   H(1-r_{ij} ), \label{eq.phi} 
\end{equation} 
where $r_{ij} = |\vect{r}_{ij}|; \vect{r}_{ij}\equiv 
\vect{r}_j - \vect{r}_i$. We also define the unit vector 
$\vect{n}_{ij} \equiv \vect{r}_{ij}/r_{ij} $ between particles 
$i$ and $j$. 
The Heaviside step function $H$ in Eq.~\ref{eq.phi} is defined by
\begin{equation}
H(x) \equiv   
  \left\{
    \begin{array}{ll}
      0           & \mbox{  for }  \ \  x <   0, \\
      \frac{1}{2} & \mbox{  at   }  \ \ x =   0, \\
      1           & \mbox{  for }  \ \  x >   0. 
    \end{array}
  \right.
\label{eq.h}
\end{equation}

Espa\~nol and Warren proposed that the following simple form of the random 
and dissipative forces, as follows\cite{95Espanol}:
\begin{eqnarray}
{\vect{F}}_{i j}^{\rm R} &=& \sigma  \omega_{\rm R} (r_{i j})  \vect{n}_{i j} 
\frac{ \zeta_{ij }}{\sqrt{\Delta t}},
\label{eq.fr} \\
{\vect{F}}_{i j}^{\rm D} &=& -\gamma \omega_{\rm D} (r_{i j})
 \left(  \vect{v}_{i j}\cdot \vect{n}_{i,j} 
 \right) \vect{n}_{i j} , \label{eq.fd}
\end{eqnarray}
where  $\zeta_{ij}$ is a Gaussian random valuable with zero mean and 
unit variance, chosen independently for each pair $(i, j)$ of interacting 
particles at each time-step and $\zeta_{ij} = \zeta_{ji}.$ The strength 
of the dissipative and random forces is determined
by the dimensionless parameter $\sigma$ and $\gamma$, respectively. 
The parameter $\Delta t$ 
is a dimensionless time-interval of integrating the equation of motion.

Now we consider the fluctuation-dissipative theorem of DPD method.
The time evolution of the distribution function of the DPD system is 
governed by Fokker-Planck equation\cite{95Espanol}. The system 
evolves to the same steady state as the Hamiltonian 
system, that is, Gibbs-Boltzmann canonical ensemble, if the coefficients 
of the diffusion and random force terms  have the following  relations:
\begin{eqnarray}
\omega_{\rm D} &=& \left( \omega_{\rm R}\right)^2, \label{eq.dis1}\\
\sigma^2  &=& 2 T \gamma , \label{eq.fluct}
\end{eqnarray} 
where  $T$ is the dimensionless equilibrium-temperature.
The forms of the weight functions $\omega_{\rm D}$ and $\omega_{\rm R}$ 
are not specified by the original DPD algorithm. We adopted  the following simple 
form as the weighting functions\cite{95Espanol}:
\begin{equation}
 \omega_{\rm R}(r) = \left\{ \omega_{\rm D} (r) \right\}^{1/2}
 = \omega\left( r\right). 
\label{eq.omegar} 
\end{equation}
Here the function $\omega$ is defined by\cite{95Espanol,97Groot}.
\begin{equation}
\omega(x) \equiv (1-x) H(1-x). \label{eqomega}
\end{equation}

\begin{table}
\begin{center}
\large
\begin{tabular}{|c|ccc|}   \hline
$a_{ij}$ & W & A & B \\ \hline
\ \ \ W \ \ \ & 25 & 0 & 50 \\
\ \ \ A \ \ \ & 0  & 25 & 30 \\
\ \ \ B \ \ \ & 50 & 30 & 25  \\ \hline
\end{tabular} 
\end{center}
\caption{\footnotesize \label{aij}The table of  the coefficients $a_{ij}$,
which depend on kinds of particles $i$ and $j$; W is a ``water" particle, 
A is a ``hydrophilic" particle and B is a ``hydrophobic" one.}
\end{table}

Finally, we use the following form as  the bond-stretching force:
\begin{equation}
\vect{F}_{i j}^{\rm B} = -\nabla_i \phi_{ij}^{\rm B},                 
\label{eq.fbond}
\end{equation}
where $\phi_{\rm B}$ is the dimensionless bond-stretching potential energy. 
When particles 
$i$ and $j$ are connected in the same molecule, they interact with each others by 
the following potential energy:
\begin{equation}
\phi_{ij}^{\rm B} \equiv \phi^{\rm B} (r_{ij}) = \frac{1}{2} a_{\rm B} 
r_{ij}^2,  \mbox{ \ \ \   if $i$ is connected to $j$.}      
\label{eq.bond}  
\end{equation}
Here $a_{\rm B} $ is the potential energy coefficient and  
$r_0$ is the equilibrium bond length.

\subsection*{\normalsize Simulation Model and Parameters}
As the model, we used modified Jury model molecule to  
dimer which is composed of
hydrophilic particle  (A) and hydrophobic one (B)\cite{99Jury}.
Water molecules are also modeled to coarse-grained particles W.
All mass of particles are assumed to unity.
The number density of particle $\rho$ is set to $\rho=6$.
The number of modeled amphiphilic molecules AB is shown as $N_{\rm AB}$,
the number of water $N_{\rm W}.$
Total number of particle $N\equiv 2N_{\rm AB} + N_{\rm W}$ 
is fixed to $N=10000.$
The simulation box is set to cubic.
The dimensionless length of the box $L$ is 
\begin{equation}
L=\left(\frac{N}{ \rho } \right)^{ \frac{1}{3}} 
\sim 11.85631.   \label{eq.L}
\end{equation}
We use periodic boundary condition in simulation.
The interaction coefficient $a_{ij} $ in Eq.~\ref{eq.phi} is given in
Table~\ref{aij}.

The coefficient of bond-stretching interaction potential $a_{\rm B}$  
is adopted as follows:
\begin{equation}
a_{\rm B} = 100 . \label{eq.abr0}
\end{equation}
We use the dimensionless time-interval of step as $\Delta t = 0.05.$

As the initial configuration, 
all particle were located randomly.
The velocity of each particle are distributed
to satisfy Maxwell distribution with dimensionless
temperature $T.$ 

The dimensionless strength of dissipative and random forces are 
$\gamma = 5.6250$ and $\sigma = 3.3541 \sqrt{T}.$

\section*{\normalsize SIMULATION RESULTS}
\subsection*{\normalsize Dynamics of Structure Formation}
\begin{figure}[h]
\begin{center}
\includegraphics[width=14cm,clip]{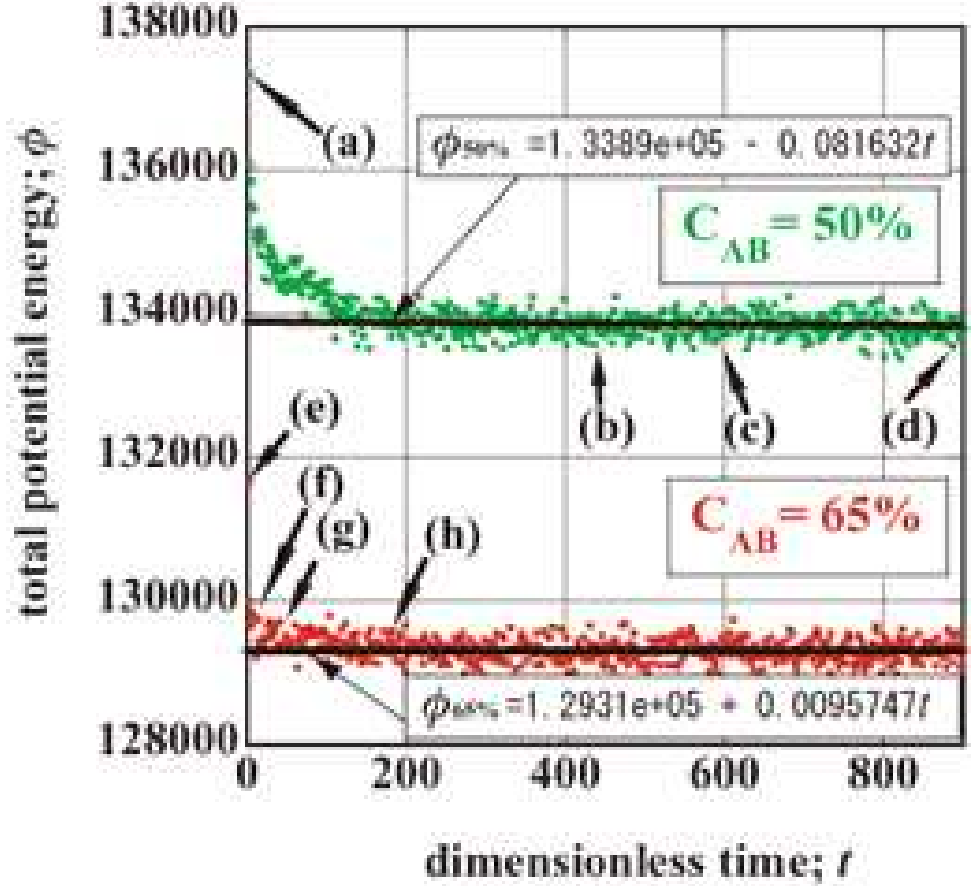}
\end{center}
\caption{\footnotesize \label{fig:energy}
Time evolution of total potential energy. 
Green and red dots denote total potential energy for 
concentration of AB dimes $c_{\rm AB} = $ 50\%, and
65 \%, respectively. 
Dimensionless temperature 
is kept to $T=0.5.$ 
We make snapshots at  eight times, i.e., 
$t=$0 (a), 450 (b), 600 (c), 900 (d) for $c_{\rm AB} = $ 50\%, 
and 
$t=$0 (e), 30 (f), 75 (g), 195 (h) for $c_{\rm AB} = $ 65\%.
The snapshots are drawn in Figs.~\ref{fig:hexal}
and \ref{fig:lamella}.
Using the least squares method, we fit data in the region of 
$t \geq 200$ to liner functions. }
\end{figure}

We plot time evolution of total potential energy $\phi$ of amphiphilic molecules
for $c_{\rm AB} = $ 50\% and 65 \%, at $T=0.5$ in Fig.~\ref{fig:energy}.

\begin{figure}[h]
\begin{center}
\includegraphics[width=14cm,clip]{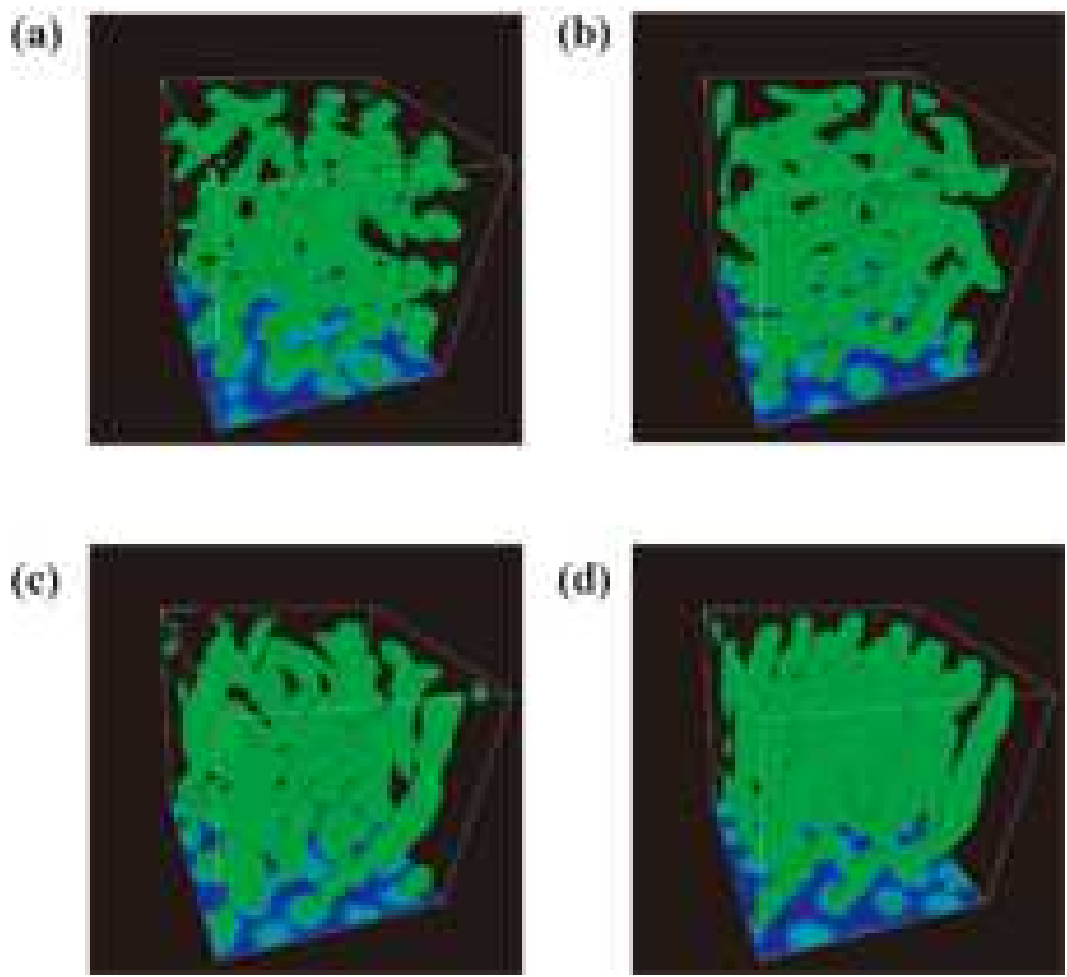}
\end{center}
\caption{\footnotesize \label{fig:hexal} 
Snapshots of time evolution of self-assembly for 
$c_{\rm AB} = $ 50\% at $T=0.5$. 
As the initial configuration(a),
AB dimers and W monomers are distributed randomly in the simulation box.
Velocities of all particles are distributed following Maxwell distribution.
To make self-assembled structure of supramolecule clearly understandable, 
we pick up surface of the   self-assembled structure, and interpolate 
the surface as if surface were continuous body.
As time passes, hexagonal structure (${\rm H}_1$) grows 
like $t=$450 (b), 600 (c), 900 (d).}
\end{figure}

For $c_{\rm AB} = $ 50\%, supramolecule structure grows to
hexagonal phase around $t=900.$ (See Fig.\ref{fig:hexal}.)
Amphiphilic molecules start to make a self-aggregation  
from random configuration (Fig.\ref{fig:hexal}(a)).
Then, local order   grows like  Fig.\ref{fig:hexal}(b).
As time goes, the local order  coalesces and the cylinder micelle structure 
grows up like  Fig.\ref{fig:hexal}(c).
At $t\approx 900$, hexagonal phase (H$_1$) was  constructed like 
Fig.\ref{fig:hexal}(d).
The total potential energy of amphiphilic molecules decreases during 
the process from (a) to (b), more than from (b) to (d). 
(See Fig.\ref{fig:energy}.)
Time dependence of the total potential energy is approximated to
$\phi_{50\%}(t) = 1.3389\times10^{5} - 0.081632\ t,$ 
by least squares method for $t \geq 200.$

\begin{figure}[h]
\begin{center}
\includegraphics[width=14cm,clip]{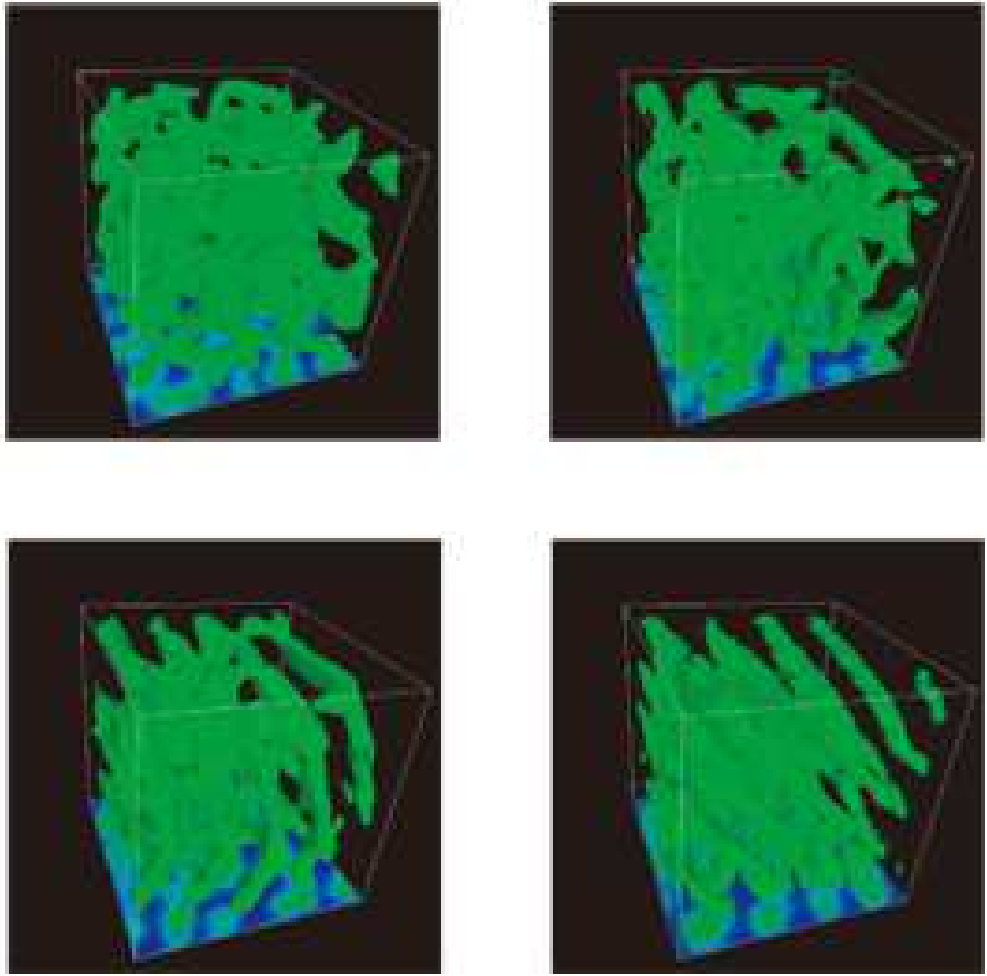}
\end{center}
\caption{\footnotesize \label{fig:lamella}
Snapshots of time evolution of self-assembly for 
$c_{\rm AB} = $ 65\% at $T=0.5$. 
Visualization method, interaction forces, system size and other physical quantities except concentration of AB dimes are the same values as Fig.~1.
As time passes, lamellar structure (${\rm L}_{\alpha}$) grows 
like $t=$0 (e), 30 (f), 75 (g), 195 (h).
The time to form the self-assembled structure is shorter than
${\rm H}_1$ phase.}
\end{figure}

For $c_{\rm AB} = $ 65\%, 
the lamellar phase $({\rm L}_{\alpha})$ was 
constructed at $t\approx 195$ like Fig.~\ref{fig:lamella}.
Time dependence of the total potential energy of amphiphilic molecules $\phi_{65\%}$
is approximated to $\phi_{65\%}(t) = 1.2931\times10^{5} - 0.0095747\ t.$ 
(See Fig.~\ref{fig:energy}.)
Comparing $\phi_{65\%}(t)$ with  $\phi_{50\%}(t)$,
it is found that, 
the lamellar phase $({\rm L}_{\alpha})$ for  
$c_{\rm AB} = 65\%$ is stable at $t=195$, 
though structure for $c_{\rm AB} = 50\%$ at $t=195$ does not
reach the stable phase, {\it i.e.}, H$_1$ phase.
In other words, it takes a longer time to form the hexagonal  phase (${\rm H}_1$) 
than the lamellar phase $({\rm L}_{\alpha})$.

\subsection*{\normalsize Phase Diagram}
\begin{figure}[h]
\begin{center}
\includegraphics[width=7cm,clip]{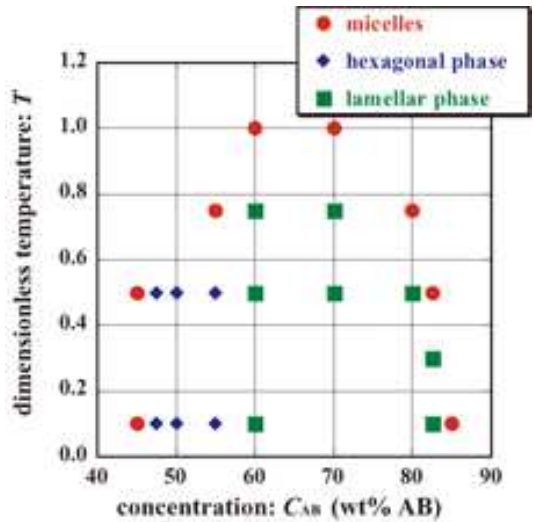}
\end{center}
\caption{\footnotesize \label{phase}
Phase diagram of  AB dimers in solution W monomers.
Horizontal axis is concentration of AB dimers $C_{\rm AB}$.
Vertical one is dimensionless temperature $T$.
Filled circles denote the micelles phase (${\rm L}_{\rm 1}$).
Hexagonal (${\rm H}_1$) and lamellar (${\rm L}_{\alpha}$) phases
are indicated by filled diamonds and squares, respectively.
This phase diagram is qualitatively consistent with the previous work by Jury
{\it et al.}\cite{99Jury}. }
\end{figure}

In the previous subsection, we showed the dynamics of 
structure formation for two cases, {\it i.e.}, 
$(c_{\rm AB}, T) = (50\%, 0.5) $ and $(65\%, 0.5)$.
Next, we simulated the other cases $(c_{\rm AB}, T )$ to
obtain phase diagram of AB dimers in W monomers in Fig.~\ref{phase}.
The phase diagram (Fig. \ref{phase}) is qualitatively consistent 
with Jury's result\cite{99Jury}.
The phase diagram (Fig. \ref{phase}) also agrees with experimental result\cite{83Mitchell}.
However, we considered temperature dependence of covariant bond interaction,
which denotes that a length of AB dimers is changeable by temperature.
From this property, our modified Jury model is more sensitive to a change of temperature 
than original Jury model where AB dimers are rigid\cite{99Jury}.

\onecolumn

\twocolumn
\section*{\normalsize RESULTS AND DISCUSSION}
We found 
that it takes a longer time to form the hexagonal  phase (${\rm H}_1$) 
than the lamellar phase $({\rm L}_{\alpha})$ where the system has higher  order 
than ${\rm H}_1$ phase.
This behavior is different from ordinary structure formation,
where  it takes a longer time to form the higher ordered phase than
to form the lower ordered phase.

We would reach the conclusion that, in the process of  structure formation in
lamellar phase, the lamellar structure grows up directly from randomly 
distributed configuration of  AB dimers. 
This behavior is contrary to our expected scenario that
hexagonal cylinder structure grows first, 
then the hexagonal cylinders coalesce and finally lamellar appears.

Concerning the simulation model,
it is easy to extend  model of amphiphilic molecules for the chain-like configuration.
We have already studied the dependence of interaction parameter of phase diagram\cite{04Nakamura}.

\section*{\small \bf  \it Acknowledgments}  
This research was partially supported 
by the Ministry of Education, Culture, Sports, Science and Technology, 
Grant-in-Aid for Scientific Research (C), 2003, No.15607019.
The DPD simulation algorithm was introduced to me by Drs. Michel Laguerre 
and Reiko ODA when the author was invited to  Institut Europen de Chimie 
et de Biologie  under CNRS Program in France.
Prof. Hajime TANAKA in the University of Tokyo introduced  to  me experiments of  
${\rm C}_{12} {\rm E}_6$.
Mr. Ryo KAWAGUCHI was helped for simulation.

\bibliographystyle{unsrt}
\small

\bibliography{reference2003}

\end{document}